\documentclass[prl,twocolumn,superscriptaddress]{revtex4}
\usepackage{epsfig}
\usepackage{times}
\usepackage{amsmath}
\usepackage{amsfonts}
\usepackage{amssymb}
\usepackage[usenames]{color}
\usepackage{graphicx}

\newcommand{\beq}{\begin{equation}}
\newcommand{\eeq}{\end{equation}}
\newcommand{\beqa}{\begin{eqnarray}}
\newcommand{\eeqa}{\end{eqnarray}}

\begin{document}
\title{Quantum Zeno effect with a superconducting qubit
}

\author{Y. Matsuzaki}
\email{matsuzaki@ASone.c.u-tokyo.ac.jp}
\affiliation{
NTT Basic Research Laboratories, NTT Corporation, 
Kanagawa, 243-0198, Japan
}
\author{S. Saito}
\affiliation{
NTT Basic Research Laboratories, NTT Corporation, 
Kanagawa, 243-0198, Japan
}
\author{K. Kakuyanagi}
\affiliation{
NTT Basic Research Laboratories, NTT Corporation, 
Kanagawa, 243-0198, Japan
}
\author{K. Semba}
\affiliation{
NTT Basic Research Laboratories, NTT Corporation, 
Kanagawa, 243-0198, Japan
}

\begin{abstract}
 Detailed schemes are investigated for experimental verification of Quantum
 Zeno effect with a superconducting qubit. A superconducting qubit is
 affected by a dephasing noise whose spectrum is $1/f$, and so the decay
 process of a
 superconducting qubit shows a  naturally
 non-exponential behavior due to an infinite correlation time of
 $1/f$ noise.
 Since projective measurements can easily influence the
 decay dynamics having such non-exponential feature, 
 a superconducting qubit is a promising system to observe Quantum
 Zeno effect.
 We have studied how a sequence of projective measurements can change
 the dephasing process and also we have
 suggested experimental ways to observe Quantum Zeno effect with a superconducting qubit.
 It would be possible to demonstrate our prediction in the current technology.
\end{abstract}

\maketitle

Quantum Zeno effect(QZE) is one of fascinating phenomena which quantum
mechanics predicts.  
A sequence of projective measurements to an unstable
system can
suppress the decay process of the state
\cite{BECG01a,Cook01a,FacchiNakazatoPascazio01a}.
This phenomena will be observed
 if the time interval of
projective measurements is sufficiently small and the decay behavior in the time
interval is quadratic.
Although it was proved that an unstable system shows a
quadratic behavior in the initial stage of the decay
\cite{NakazatoNamikiPascazio01a}
, it is
difficult to observe such quadratic decay behavior
experimentally, because the time region to show such quadratic behavior
is usually much shorter than typical time resolution of a measurement
apparatus in the current technology.
After
showing the quadratic decay, unstable system shows an exponential decay
\cite{NakazatoNamikiPascazio01a}
and QZE doesn't occur through
projective measurements to a system which decays exponentially.
Due to such difficulty, in spite of the many effort to observe the QZE, there was
only one experimental demonstration to suppress the decay process of an
unstable state
\cite{FischerGutierrezRaizen01a}.
Note that, except this experiment, all previous
demonstration of QZE didn't focus on a decoherence process
caused by a coupling with environment but focused on a suppression of a unitary
evolution having a finite Poincare time such as Rabi
oscillation
\cite{
BalzerHuesmannNeuhauserToschek01a,MolhaveDrewsen01a,NagelsHermansChapovsky01a,
StreedMunBoydGretchenCampbellMedleyKetterlePritchard01a,ToschekWunderlich01a}.
Such approach to change the behavior of the unitary evolution by
measurements are experimentally easy to be demonstrated, but is different from the original
suggestion of QZE for the decay process of unstable systems
\cite{BECG01a,Cook01a,FacchiNakazatoPascazio01a}
with a decoherence process.
  Throughout this paper, we consider only such QZE to change
  decoherence behavior.
  
In this paper, we suggest a way to demonstrate QZE for
the decay process of unstable system experimentally with a superconducting qubit.
A superconducting qubit is one of candidates to realize quantum information
processing and, for a superconducting qubit, the quadratic decay
has been observed in an experiment
\cite{YoshiharaHarrabiNiskanenNakamura01a,KakuyanagiMenoSaitoNakanoSembaTakayanagiDeppeShnirman01a}
, which is
necessary condition to observe QZE experimentally.
Moreover, a high fidelity single qubit measurement has already
been constructed in the current technology
\cite{ClarkeWilhelm01a}.
A superconducting flux  qubit has been traditionally measured by
superconducting quantum interference device(SQUID)
\cite{clarke2004squid}.
The state of a SQUID
is switched from zero-voltage state to a finite voltage state for a
particular quantum state of the qubit, while no switching
occurs for the other state.
Such switching transition produces a macroscopic signal to
construct a measurement for a superconducting flux qubit.
Also, entirely-new qubit readout method such as JBA(:Josephson
Bifurcation Amplifier) has been demonstrated
\cite{SiddiqiVijayPierreWilsonMetcalfeRigettiFrunzioDevoret01a,LupascuSaitoPicotGrootHarmansMooij01a}.
The JBA has
advantages in its readout speed, high sensitivity, low backaction
\cite{LupascuSaitoPicotGrootHarmansMooij01a}
and absence of on-chip dissipative process. It is also studied JBA
readout mechanism
\cite{NakanoSaitoSembaTakayanagi01a}
and the projection conditions
\cite{KakuyanagiNakanoKageiKoibuchiSaitoSemba01a}
of the
superposition state of a qubit. All these properties are
prerequisite in observing the QZE.
So a superconducting qubit is a promising system to verify
QZE for an unstable state.

We study a general decay process of unstable system. 
Although a decay behavior of unstable system has been studied and conditions for quadratic decay have been shown by several
authors\cite{NakazatoNamikiPascazio01a,MakhlinShnirman01a,MartinisNamAumentadoLang01a,Schulman01a}
,
we introduce a simpler solvable model
and we confirm the conditions for
 the exponential decay and the
quadratic decay, respectively.
Also, from the analytical solution of the model, we derive a master
equation for $1/f$ noise.
We consider an interaction Hamiltonian to denote a coupling with
an environment such as $H_I=\lambda f(t)\hat{A}$
where $f(t)$ is a classical normalized Gaussian noise, $\hat{A}$ is an
operator of the system, and $\lambda $ denote a coupling constant.
Also, we assume non-biased noise and therefore  $\overline{f(t)}=0$
is satisfied where this over-line denotes the average over the ensemble of
the noises.
In an interaction picture, by solving the Schrodinger equation and
taking the average over the ensembles,
we
obtain
           \begin{eqnarray}
       &&\rho_I(t) -\rho _0
        =\sum_{n=1}^{\infty
  }(-i\lambda )^n\int_{0}^{t}dt_1\int_{0}^{t_1}dt_2\cdots
  \int_{0}^{t_{n-1}}dt_n\nonumber \\
  &&\overline{f(t_1)f(t_2)\cdots f(t_n)}[\hat{A},[\hat{A},\cdots ,[\hat{A},\rho_0
  ],\cdots ]]\label{analytical} 
     \end{eqnarray}
     where $\rho _0=|\psi \rangle \langle \psi |$ is an initial state
     and $\rho _I(t)$ is a state in the interaction picture.
     Throughout this paper, we restrict ourself to a case that     
     the system Hamiltonian commutes with the operator of $1/f$ noise as $[H_s, \hat{A}]=0$.
     Firstly, we consider a case that the correlation time of the noise $\tau
     _c\equiv \int_{0}^{\infty }\overline{f(t)f(0)}dt$ is
much shorter than the time resolution of experimental
apparatus, which is valid condition for the most of unstable
     systems. 
Since the correlation time of the noise is short, we obtain 
$\int_{0}^{t}\int_{0}^{t'}\overline{f(t')f(t'')}dt'dt''=\int_{0}^{t}d\tau
(t-\tau )\overline{f(\tau )f(0)}\simeq t\tau _c$. 
Also, since the noise $f(t)$ is Gaussian,
$\overline{f(t_1)f(t_2)\cdots f(t_n)}$ can be decomposed of a product of
two-point correlation $\overline{f(t_i)f(t_j)}$,
and so we obtain
\begin{eqnarray}
 \rho _I(t)\simeq \sum_{A,A',\nu ,\nu '}|A\nu \rangle \langle A\nu |\rho
  _0|A'\nu '\rangle \langle A'\nu '|e^{-\lambda ^2\tau _c|A-A'|^2t}\ \ 
\end{eqnarray}
where $|A\nu \rangle $ is an eigenstate of the operator $\hat{A}$ and
  $\nu $ denote a degeneracy.
So 
a dynamical fidelity $ F\equiv \langle \psi |e^{iH_st}\rho(t) e^{-iH_st}
  |\psi \rangle $, a distance between the
  state
  $\rho (t)$ and a state $e^{-iH_st}\rho _0e^{iH_st}$,
  becomes a sum of exponential decays.
\begin{eqnarray}
F\simeq  \sum_{A,A',\nu ,\nu '}
 |\langle A\nu |\psi \rangle |^2 |\langle A'\nu '|\psi \rangle |^2
 e^{-\lambda ^2\tau _c|A-A'|^2t}
\end{eqnarray}
Secondly, when the correlation time of the noise is much longer than the time resolution of the
apparatus such as $1/f$ noise having an infite correlation time,
we obtain
$\int_{0}^{t}\int_{0}^{t'}\overline{f(t')f(t'')}dt'dt''\simeq \frac{1}{2}t^2$.
    So, by taking average over the
ensemble of noise in (\ref{analytical}), we obtain
\begin{eqnarray}
 \rho _I(t)\simeq \sum_{A,A',\nu ,\nu '}|A\nu \rangle \langle A\nu |\rho
  _0|A'\nu '\rangle \langle A'\nu '|e^{-\frac{1}{2}\lambda ^2|A-A'|^2t^2}\ \ \label{analytic-1-f}
\end{eqnarray}
So we obtain a master equation for $1/f$
noise as
$\frac{d\rho _I(t)}{dt}= -\lambda ^2 t[\hat{A},[\hat{A},\rho _I(t)]]\label{master-equation}$.
The behavior of the dynamical fidelity
becomes quadratic in the early stage of the decay ($t\ll
\frac{1}{\lambda }$) as
\begin{eqnarray}
F&\simeq &\sum_{A,A',\nu ,\nu '}
 |\langle A\nu |\psi \rangle |^2 |\langle A'\nu '|\psi \rangle |^2
 e^{-\frac{1}{2}\lambda ^2|A-A'|^2t^2}
 \nonumber \\
 &\simeq& 1-\frac{1}{2}\lambda ^2t^2 \sum_{A,A',\nu ,\nu '}|A-A'|^2
 |\langle A\nu |\psi \rangle |^2 |\langle A'\nu '|\psi \rangle |^2
 \ \ \ \ \ \ 
\end{eqnarray}
These results show that an unstable system has an exponential decay for $t\gg \tau
_c$, while 
a quadratic
decay occurs for $t\ll \tau _c$.

Let us summarize the QZE. Usually, to observe QZE, survival probability is
chosen as a measure for the decay. However, we use a dynamical fidelity
to observe the QZE rather than a survival probability to take into account
of the effect of a system Hamiltonian.
We consider
a sequence of projective
measurements $\hat{\mathcal{P}}(k )=e^{-iH_s k\tau }
|\psi\rangle\langle\psi| e^{i H_s k\tau }$ with $\tau =\frac{t}{N}$ and
$k=1,2,\dots,N$
to an unstable state
where
$N$ denotes the
number of the measurements performed during the time
$t$.
For noises whose correlation time is short,
a dynamical fidelity without
measurements becomes a sum of exponential decay terms such as
$F(t)=\sum_{j=1}^{m}c_je^{-\Gamma _j
t}$. The success probability to
project the unstable state into the target states
becomes
$P(N)=\big{(}\sum_{j=1}^{m}c_je^{-\Gamma _j
\tau }
\big{)}^N\simeq 1-t\sum_{j=1}^{m}c_j\Gamma _j $, 
and so the success probability decreases linearly as the time
increases.
On the other hand, if the
dynamical fidelity has a quadratic decay without projective
measurements such as $F=e^{-\Gamma ^2t^2}$, we obtain the success probability to
project the unstable state into the state $e^{-iH_st}|\psi \rangle $
becomes as following.
$P(N)=\Big{(}1-\Gamma ^2\tau ^2+O(\tau ^4)\Big{)}^N \simeq 1-\Gamma ^2\frac{t^2}{N} $.
So, by increasing the number of the measurements, the success
probability goes to unity, and this means that the time evolution of
this state is confined into $e^{-iH_st}|\psi \rangle $ which is a purely
unitary evolution without noises, and so one can observe the QZE. 

It is known that a superconducting qubit is mainly affected by two decoherence
sources, a dephasing
whose spectrum is $1/f$ and a relaxation whose spectrum is white. The $1/f$ noise causes a
quadratic decay to the quantum states as we have shown. Moreover, 
such quadratic decay has already been observed experimentally
\cite{YoshiharaHarrabiNiskanenNakamura01a,KakuyanagiMenoSaitoNakanoSembaTakayanagiDeppeShnirman01a}
. On the
other hand, since the relaxation process  from an excited state
$|1\rangle $
to a ground state $|0\rangle $
where a high frequency is cut off, 
the correlation time of the environment is extremely small
and so only
an exponential decay can be observed for a relaxation process in the current technology.
Therefore, when the dephasing is relevant and the relaxation is negligible, it should be
possible to observe QZE with a
superconducting qubit as following.
    \begin{figure}[h]
     \begin{center}
      \includegraphics[width=4.7cm]{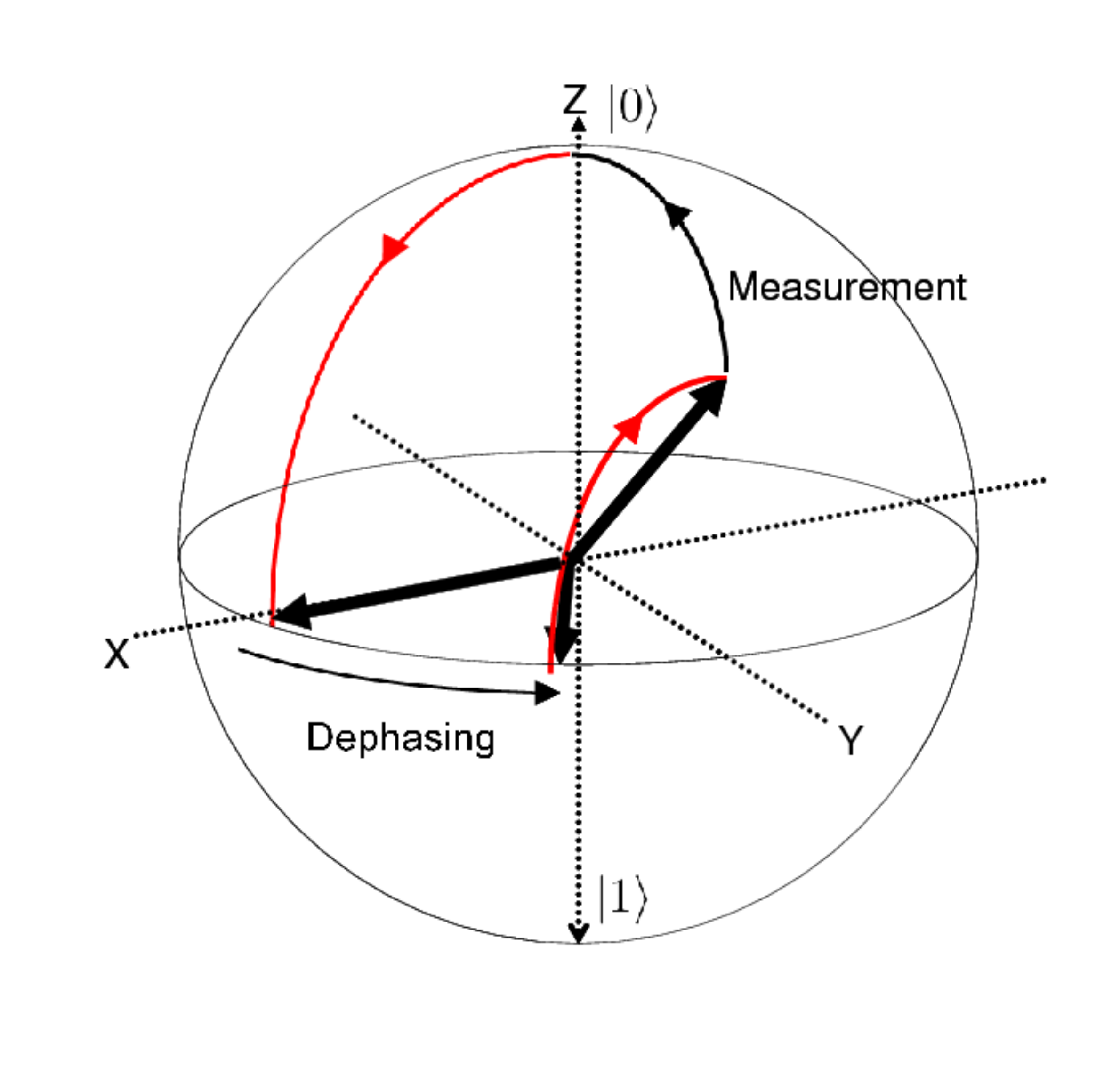}
     \end{center}
     \caption{A schematic of quantum states in a Bloch sphere to show how QZE is observed with a
     superconducting qubit. An initial state is prepared in $|+\rangle $,
     and the state has an unknown rotation around $z$ axis due to a
     dephasing. To construct a measurement $|+\rangle \langle
     +|$, one performs a $\frac{\pi }{2}$ rotation $U_y$ around $y$ axis, performs a
     measurement $|0\rangle \langle 0|$, and performs a
     $\frac{\pi}{2}$ rotation
     $U_{y}^{\dagger}$. If a measurement interval is much smaller than a
     dephasing time, this measurement $|+\rangle \langle +|$
     recovers a state into the initial state with almost unity success
     probability.
     }
      \label{bloch}
    \end{figure}
Firstly, one prepares an initial
state 
$|+\rangle =\frac{1}{\sqrt{2}}|0\rangle +\frac{1}{\sqrt{2}}
|1\rangle $ which is an eigenstate of $\hat{\sigma }_x$.
Secondly, in a time interval $\tau =\frac{t}{N}$,
one continues to perform projective
measurements $|+\rangle \langle
+|=\frac{1}{2}(\hat{\openone}+\hat{\sigma }_x)$ to the
superconducting qubit where $N$ is the number of the measurement
performed.
For simplisity, let us make an assumption that an effect of a system Hamiltonian is negligible
compared with the dephasing effect. (Since this assumption could be
unrealistic for a superconducting qubit, we will relax this condtion and discuss more rigorous case later.)
 Note that we perform a selective measurement here to consider only
a case to project the state into $|+\rangle \langle
+|$ and, if the state is projected into the other state,
we discard the state as a failure case. 
As a result, due to the quadratic decay behavior caused by $1/f$ noise, the success
probability to project the state into a target state $N$ times goes to
a unity as the number of the measurements becomes larger, and therefore
one can observe QZE(Fig.\ref{bloch}).
Note
that a direct measurement of
$\hat{\sigma }_x$ with a superconducting qubit has not been constructed
yet experimentally.  So, in order to know a measurement result of $\hat{\sigma
}_x$  in the current
technology, one has to perform a $\frac{\pi }{2}$ rotation around $y$ axis before and after performing a projective
 measurement about $\hat{\sigma }_z$.
  However, recently, a coupling
  about $\hat{\sigma }_x$ between a superconducting qubit and a flux bias control line
 has been demonstrated
 \cite{fedorov2010strong,Zhu01a}, which shows a possibility to
  realize a direct measurement of $\hat{\sigma }_x$ in the near future. Since it is not
  necessary to perform preliminary rotations around $y$ axis, this direct measurement of
  $\hat{\sigma }_x$ has advantage in its readout speed.

\begin{figure}[h]
   \begin{center}
    \includegraphics[width=5.9cm]{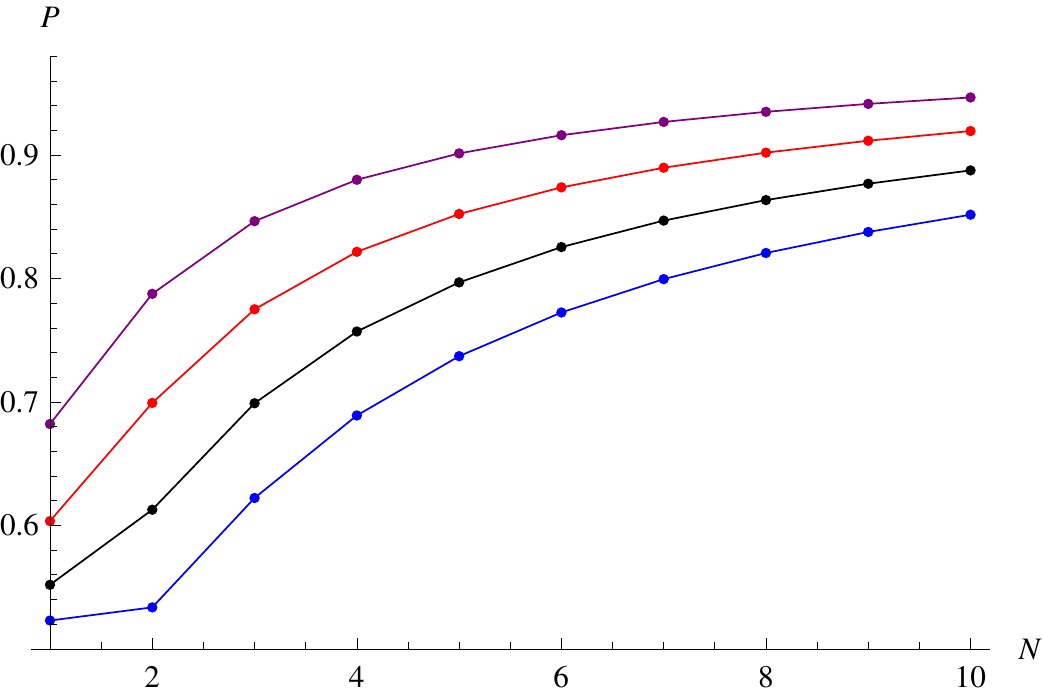}
   \end{center}
   \caption{A success probability to perform projective measurements into
   a target state under the effect of dephasing and relaxation is plotted.
   The horizontal axis and the vertical axis denote the success probability
   and the number of measurements respectively
   The lowest line
   is
   for $t=35$(ns), and the other lines
   are for
   $t=30,25,20$(ns), respectively. As one increases the
   number of measurements,
   the success probability increases.
  Here, we assume a relaxation time $T_1=1\ \mu $s and a dephasing time
   $T_2=20$ ns, respectively.}
    \label{successprobability}
  \end{figure}

In the above discussion, the effect of the relaxation and
system Hamiltonian
is not
 taken into account. Since
 they are not always negligible in a superconducting qubit, it
 is necessary to investigate whether one can observe QZE or not under
 the influence of them.
 When considering the effect of dephasing and relaxation whose spectrum are $1/f$ and
 white respectively, we use a master equation as
 following
 \begin{eqnarray}
  &&\frac{d\rho _I(t)}{dt}=-\frac{1}{2}\Gamma _1\Big{(}\hat{\sigma }_+\hat{\sigma }_-\rho_I(t) +\rho_I(t)
 \hat{\sigma }_+\hat{\sigma }_-\nonumber \\
 &&-2\hat{\sigma }_- \rho _I(t)\hat{\sigma }_+\Big{)}
  -\frac{1}{2}(\Gamma _2) ^2t
   [\hat{\sigma }_z, [\hat{\sigma }_z,\rho
   _I(t)]] 
\end{eqnarray}
where $\Gamma _1$ and $\Gamma _2$ denote a decoherence rate of relaxation
and dephasing respectively. In this master equation, the first part is a
Lindblad type master equation to denote a relaxation, and the
second part denotes a dephasing whose spectrum is $1/f$ coming from the
fluctuation of $\epsilon $.
Also, we assume that a system Hamiltonian
is $H_s=\frac{1}{2}\epsilon \hat{\sigma }_z+\frac{1}{2}\Delta 
\hat{\sigma } _x\simeq \frac{1}{2}\epsilon \hat{\sigma }_z$ for
$\epsilon \gg \Delta $, because we
have derived a master equation for $1/f$ noise only when the system
Hamiltonian commutes with the noise operator of $1/f$ fluctuation.
We find an analytical solution of this equation, and when the initial state
is $|\psi \rangle =|+\rangle$, we obtain
\begin{eqnarray}
 \rho
  (t)=e^{-iH_{s}t}\Big{(}\frac{1}{2}e^{-\Gamma
  _1t} |1\rangle \langle |1+\frac{1}{2}e^{-\frac{1}{2}\Gamma _1t-
(\Gamma
  _2)^2 
  t^2}|0\rangle \langle 1|\nonumber \\
  +\frac{1}{2}e^{-\frac{1}{2}\Gamma _1t-
   (\Gamma
  _2)^2
  t^2}|1\rangle \langle
  0|+(1-\frac{1}{2}e^{-\Gamma _1t})
|0\rangle \langle 0|
\Big{)}e^{iH_{s}t}\ \ \label{analytical-solution-master} 
\end{eqnarray}
Note that,
while the $1/f$ noise causes a
quadratic dephasing, the relaxation causes an
exponential decay, which cannot be suppressed
by projective measurements.
Here, we consider the effect of system Hamiltonian, and
so we
perform a projective measurement to the state $e^{-iH_st}|+ \rangle $.
Since there always exists a time-dependent single qubit rotation $U_t$ to satisfy
$U_te^{-iH_st}|+ \rangle  =|0\rangle$,
this measurement can be realized by performing the single qubit rotation before and
after a measurement of $\hat{\sigma }_z$
Note that
this single qubit rotation $U_t$ can be performed in a few ns by using a
resonant microwave\cite{kutsuzawa-rotation}. In this
paper, we call
the entire process including $U_t$ as ``measurement'' for simplicity.
The success probability $P(N)$ to project the state into the target state
is
calculated as
 \begin{eqnarray}
 P(N) =(\frac{1}{2}+\frac{1}{2}\cdot
  e^{-\frac{\tau}{2T_1}-\frac{\tau^2}{(T_2)^2}})^N
 \end{eqnarray}
where $T_1=(\Gamma _1)^{-1}$ and $T _2=(\Gamma_2)^{-1}$ denote a
relaxation time and a dephasing time respectively.
So we obtain
$ P(N)=(\frac{1}{2}e^{\frac{\tau ^2}{2(T_2)^2}}+
 \frac{1}{2}e^{-\frac{\tau}{2T_1}-\frac{\tau^2}{2(T_2)^2}})^Ne^{-\frac{t^2}{2N(T_2)^2}}
\simeq (1-\frac{t }{4T_1})
   e^{-\frac{t^2}{2N(T_2)^2}}$
for $\frac{t}{T_1},\frac{t}{T_2}\ll 1$.
So, as long as the $T_1$ is much
larger
than $T_2$, one
can observe that
the success probability increases
as one
increases the number of the projective measurements(see Fig.\ref{successprobability}).
Note that we assume a Hamiltonian as $H_s\simeq
\frac{1}{2}\epsilon \hat{\sigma }_z$,
far from the optimal point for a superconducting
flux qubit, and so
a coherence time $T_2$ of this qubit becomes as small as tens of ns.
In the current technology, it takes tens of ns to perform JBA\cite{LupascuSaitoPicotGrootHarmansMooij01a} and so
one has to use a switching measurement to utilize a SQUID to be
performed in a few ns. The state of a SQUID remains a zero-voltage
when the state of a qubit is $|0\rangle $, while a SQUID makes a
transition to a finite voltage state to produce a macroscopic signal for
$|1\rangle$. 
One of the problems
of the SQUID measurements
is that
a transition to a finite voltage state
destroys quantum states
of the qubit
and following measurements
are not possible
after the transition.
However, as long as the state
is $|0\rangle $, the state of
a SQUID remains a zero-voltage state and so sequential measurements are
possible. Since one postselects a case that all measurement
results are $|0\rangle $ while one discards the other case as a failure,
the SQUID can be
utilized to observe QZE with the selective measurements.

Importantly, it is also possible to observe QZE at the optimal point where
$T_2$ can be as large as $\mu$s.
 A recent demonstration of coupling
 about $\hat{\sigma }_x$ between a superconducting qubit and a flux bias control line
 shows a possibility to
 have a relevant $1/f$ noise caused by a fluctuation of $\Delta$ due to
 a replacement of a Josephson junction with a SQUID
 \cite{fedorov2010strong,Zhu01a}, and the noise operator from the $1/f$ fluctuation
 becomes $\hat{\sigma }_x$ to commute a system Hamiltonian at the
 optimal point as $H=\Delta \hat{\sigma }_x$.
 So, by replacing
 the notation
 from $\hat{\sigma } _z$
 to $\hat{\sigma
 }_x$ and from $|+\rangle $ to $|0\rangle $,
 one can apply
 our analysis in this paper to a case observing QZE at the
 optimal point. (For example, in this replaced notation, an
 initial state should be
 prepared in $|0\rangle $ and frequent measurements in the $zy$ plane will
 be performed.) 
 Moreover, since $T_2$ at the optimal point is much longer than a necessary time to
 perform JBA, a sequence of measurements is
 possible for all measurement results. This motivates us to study a
 verification of QZE without postselection as following. 

\begin{figure}[h]
    \begin{center}
     \includegraphics[width=7cm]{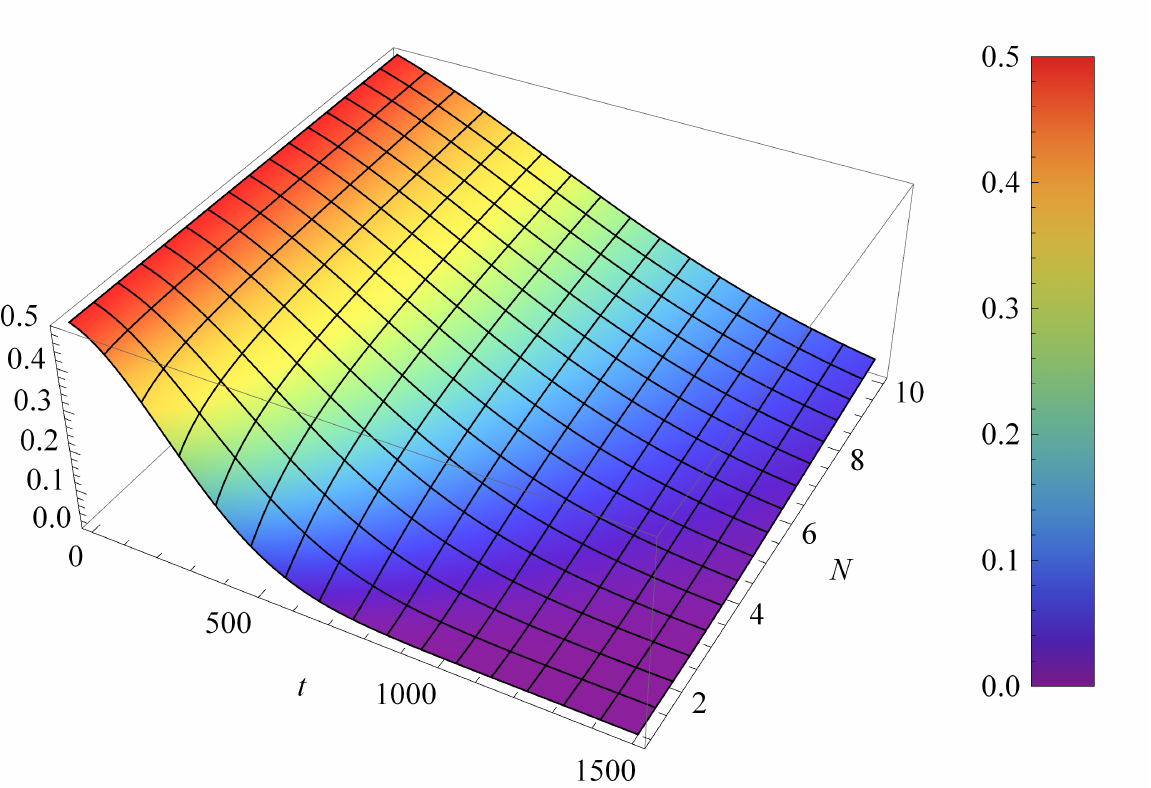}
    \end{center}
    \caption{A
    behavior of a phase term $|\langle 0|\rho
    |1\rangle |$ under the effect of non-selective measurements realized
by the JBA
    is
    plotted. This shows a suppression of the decay by measurements. Here $t$ and $N$ denote the time(ns) and the number of
    measurements, respectively.
    We assume a relaxation time $T_1=1\ \mu $s and a dephasing time
    $T_2=400$ ns, respectively. These conditions can be realized at the
    optimal point where the sytem Hamiltonian is $H_s=\Delta \hat{\sigma
    }_x$ and $\Delta $ has a $1/f$ fluctuation due to
 a replacement of a Josephson junction with a SQUID.
    }
     \label{jba}
   \end{figure}

Finally, we discuss how to observe QZE without
postselection of measurement results, which can be realized by JBA.
 We perform frequent non-selective measurements in the $xy$ plane to the
 state which was initially prepared in $|+\rangle $.
Such non-selective measurements to a single qubit is modeled as
$\hat{\mathcal{E}}(\rho )= |\phi _{+}\rangle \langle \phi _{+}|\rho |\phi _{+}\rangle \langle
   \phi _{+}| +|\phi _{-}\rangle \langle \phi _{-}|\rho |\phi _{-}\rangle
   \langle \phi _{-}|$ where
$|\phi _{+}\rangle =e^{-iH_st}|+\rangle $ and $|\phi _{-}\rangle
 =e^{-iH_st}|-\rangle $ are
 orthogonal with each other.
So, when performing this non-selective measurement with a time interval $\tau
=\frac{t}{N}$ under the influence of dephasing and relaxation, we obtain
\begin{eqnarray}
 &&\rho (N,t)=e^{-iH_st}\Big{(}\frac{1}{2}|0\rangle \langle
  0|+\frac{1}{2}e^{-\frac{t}{2T_1}-\frac{t^2}{N(T_2)^2}}|0\rangle
  \langle 1|\nonumber \\
  &+&\frac{1}{2}e^{-\frac{t}{2T_1}-\frac{t^2}{N(T_2)^2}}|1\rangle
  \langle 0|+\frac{1}{2}|1\rangle \langle 1|\Big{)}e^{iH_st}
\end{eqnarray}
where we use a result in (\ref{analytical-solution-master}).
   Since we consider a state just after performing a
measurement in the xy plane (not along z axis), the population
of a ground state becomes equivalent as the population of
an excited state.
Note that a non-diagonal term is decayed by the white noise and $1/f$
noise, and only the decay from $1/f$ noise is suppressed by the
measurements.
In Fig. \ref{jba}, we show this decay behavior of the
non-diagonal term.
A possible experimental way to remove out the effect of the white noise is measuring $\langle 0|\rho (N,t)|1\rangle $ and
$\langle 0|\rho (1,t)|1\rangle $ separately by performing a tomography,
and plotting the value of $\langle 0|\rho (N,t)|1\rangle /\langle 0|\rho
(1,t)|1\rangle = e^{-\frac{t^2}{N(T_2)^2}}$ for a fixed time $t$.
As a result one can observe the suppression of the dephasing caused by
$1/f$ noise through measurements.

In conclusion, we have studied detailed schemes for experimental
verification of QZE to a decay process with a superconducting qubit. Since a
superconducting qubit is affected by
the dephasing with a $1/f$ spectrum,
the dynamics show a quadratic decay
which is
suitable for
an experimental demonstration for QZE, while the relaxation process
has
an exponential decay to cause unwanted noise for QZE.
We have suggested a way to
observe QZE even under an
influence of relaxation.
Our prediction is feasible
in the current technology.
Authors thank H. Nakano and
S. Pascazio for valuable discussions on QZE. 
This work was done during Y. Matsuzaki's short stay at NTT corporation
and was also supported in part by Funding Program for World-Leading
Innovative R\&D on Science and Technology(FIRST), Scientific Research of
Specially Promoted Research 18001002 by MEXT, and Grant-in-Aid for
Scientific Research (A) 22241025 by JSPS.


\end{document}